\documentclass{ws-ijmpc}
\begin{document}

\markboth{Pan, Xu, Zhang, Jiang} {Lattice Boltzmann Approach to
High-Speed Compressible Flows}

\catchline{}{}{}{}{}

\title{Lattice Boltzmann Approach to High-Speed Compressible Flows}
\author{X. F. Pan, Aiguo Xu\footnote{Corresponding author}, Guangcai Zhang, Song Jiang }
\address{
National Key Laboratory of Computational Physics,  Institute of
Applied Physics and Computational Mathematics, P.O.Box
8009-26, Beijing 100088, P.R.China\\
E-mail: Xu\_Aiguo@iapcm.ac.cn}

\maketitle

\begin{history}
\received{23 April 2007} \revised{4 June 2007}
\end{history}

\begin{abstract}
We present an improved lattice Boltzmann model for high-speed
compressible flows. The model is composed of a discrete-velocity
model by Kataoka and Tsutahara [Phys. Rev. E \textbf{69}, 056702
(2004)] and an appropriate finite-difference scheme combined with an
additional dissipation term. With the dissipation term parameters in
the model can be flexibly chosen so that the von Neumann stability
condition is satisfied. The influence of the various model
parameters on the numerical stability is analyzed and some reference
values of parameter are suggested. The new scheme works for both
subsonic and supersonic flows with a Mach number up to $30$ (or
higher), which is validated by well-known benchmark tests.
Simulations on Riemann problems with very high ratios ($1000:1$) of
pressure and density also show good accuracy and stability.
Successful recovering of regular and double Mach shock reflections
shows the potential application of the lattice Boltzmann model to
fluid systems where non-equilibrium processes are intrinsic. The new
scheme for stability can be easily extended to other lattice
Boltzmann models.


\end{abstract}

\section{Introduction}

High-speed compressible flow with shocks plays an important role in
various fields, such as explosion physics, aeronautics, etc.
Efficient simulation of such a system is interesting and
challenging. The traditional method is based on a set of macroscopic
Euler equations resolved by the Finite-element or Finite-volume
schemes, where the artificial viscosity is applied or the Riemann
solver is used to capture the shock\cite{1,KXuJCP1998,YuXJ}.
According to the gas kinetic theory, a set of Euler equations
describes a system being at equilibrium. For a system with shocks,
the non-equilibrium behavior is intrinsic, so a scheme based on the
fundamental kinetic theory is to be preferred. As a new approach to
fluid dynamics, the lattice Boltzmann (LB) method\cite{lbe-1} solves
the fully discrete Boltzmann equation by using an appropriate
difference scheme
to the temporal and spatial derivatives of the distribution function $f_{i}(%
\mathbf{x},t)$, where $\mathbf{x}$ and $t$ are the position and time,
respectively, and the index $i$ corresponds to the $i$-th discrete velocity.
It recovers the desired macroscopic equations in the hydrodynamic limit and
has the potential to fill the gap between continuum description and
molecular dynamics\cite{HSPRL2006}. Besides the traditional LB originating
from the lattice gas cellular automata\cite{Swift96,noi,Xuepl,xsb2006,Karlin}%
, other versions such as finite-difference(FD)\cite%
{SetaJSP,Guo,Watari,Kataoka_NSE,Kataoka_Euler}, finite-volume(FV)\cite{FVLBM}%
, and finite-element(FE)\cite{FELBM}, etc have also been developed under the
same framework. Among these works, developing LB models for high-speed
compressible flows has long been attempted by different authors\cite%
{Kataoka_Euler,CLB}. Among the existing models for two-dimesnional
compressible fluids, the one by Kataoka and
Tsutahara(KT)\cite{Kataoka_Euler} has a simple and rigorous
theoretical background. It takes flexible ratio of specific-heat and
is superior in computational efficiency because the total number of
its discrete velocity is reduced to $9$. But similar to previous LB
models\cite{Yong2003}, the numerical stability problem remains one
of the few blocks for its practical simulation to high-Mach-number
compressible flows. In this paper we present a new scheme based on
the original discrete-velocity-model (DVM) by KT and an appropriate
finite-difference scheme combined an additional dissipation term.
With the new scheme fluid systems with high-Mach-number and/or high
ratios of pressure and density can be successfully simulated.

This paper is organized as follows. In section 2 the original
discrete-velocity-model by KT is briefly reviewed and an alternative
FD scheme is proposed for later analysis and using. A von Neumann
stability analysis is performed in section 3, from which solutions
to improve the numerical stability can be found. Several benchmark
tests are used to validate the proposed scheme in section 4. Section
5 concludes the present paper.

\section{Description of the DVM and FD scheme}

The LB equation with the Bhatanger-Gross-Krook approximation\cite{BGK} reads,

\begin{equation}
\frac{\partial f_{i}}{\partial t}+v_{i\alpha }\frac{\partial f_{i}}{\partial
x_{\alpha }}=\frac{1}{\tau }\left[ f_{i}^{eq}-f_{i}\right] \text{,}
\label{SEC1-e1}
\end{equation}%
where $f_{i}^{eq}$ is the discrete version of the local equilibrium
distribution function; $\tau $ the relaxation time; index $\alpha =1$, $2$, $%
3$ corresponding to $x$, $y$, and $z$, respectively; and $v_{i}$ the $i$-th
discrete velocity, $i=0$, $...$, $N-1$; $N$ is the total number of the
discrete velocity. Under the hydrodynamic limit the LB equation is required
to describe the following Euler equations,
\begin{eqnarray}
\frac{\partial \rho }{\partial t}+\frac{\partial (\rho u_{\alpha })}{%
\partial x_{\alpha }} &=&0,  \nonumber \\
\frac{\partial (\rho u_{\alpha })}{\partial t}+\frac{\partial (\rho
u_{\alpha }u_{\beta })}{\partial x_{\beta }}+\frac{\partial P}{\partial
x_{\alpha }} &=&0,  \label{e2} \\
\frac{\partial \rho (bRT+u_{\alpha }^{2})}{\partial t}+\frac{\partial \rho
u_{\alpha }(bRT+u_{\beta }^{2})+2Pu_{\alpha }}{\partial x_{\beta }} &=&0%
\text{,}  \nonumber
\end{eqnarray}%
where $\rho $, $u$, $T$, $P$ ($=\rho RT$) are the hydrodynamic density, flow
velocity, temperature and pressure, respectively, and $R$ is the specific
gas constant, $b$ relates to the specific-heat ratio $\gamma $ as follows, $%
b=2/(\gamma -1)$. The following constraints are imposed on the moments of $%
f_{i}^{eq}$ and $f_{i}$,
\begin{equation}
\rho =\sum_{i=0}^{N-1}f_{i}^{eq}=\sum_{i=0}^{N-1}f_{i}\text{,}  \label{e4}
\end{equation}%
\begin{equation}
\rho u_{\alpha }=\sum_{i=0}^{N-1}f_{i}^{eq}v_{i\alpha
}=\sum_{i=0}^{N-1}f_{i}v_{i\alpha }\text{,}  \label{e5}
\end{equation}%
\begin{equation}
\rho (bRT+u_{\alpha }^{2})=\sum_{i=0}^{N-1}f_{i}^{eq}(v_{i\alpha }^{2}+\eta
_{i}^{2})=\sum_{i=0}^{N-1}f_{i}(v_{i\alpha }^{2}+\eta _{i}^{2})\text{,}
\label{e7}
\end{equation}%
\begin{equation}
P\delta _{\alpha \beta }+\rho u_{\alpha }u_{\beta
}=\sum_{i=0}^{N-1}f_{i}^{eq}v_{i\alpha }v_{i\beta }\text{,}  \label{e6}
\end{equation}%
\begin{equation}
\rho \lbrack (b+2)RT+u_{\beta }^{2})u_{\alpha
}=\sum_{i=0}^{N-1}f_{i}^{eq}(v_{i\alpha }^{2}+\eta _{i}^{2})v_{i\alpha }%
\text{,}  \label{e8}
\end{equation}%
where $\eta _{i}$ is another variable introduced to make
specific-heat ratio flexible\footnote{In a practical system, the
ratio $\gamma$ provides information on the internal degrees of
freedom of molecules. For example, $\gamma$ has a certain well-known
value for an ideal, monatomic gas (like helium), and is different
for diatomic molecules like those that make up most of the
atmosphere. To formulate the DVM, the discretization and
contribution of the internal degrees of freedoms of the molecules
are represented by the constraints \eqref{e7} and \eqref{e8}. }.

Equation \eqref{SEC1-e1} may be written in non-dimensional form by using a
characteristic flow length scale $L$, reference speed $e_{r}$ and density $%
\rho _{r}$. Two reference time scales are used, $t_{c}$ to represent the
time between particle collisions and $L/e_{r}$ to present a characteristic
flow time. The resulting non-dimensional equation is
\begin{equation}
\frac{\partial \hat{f}_{i}}{\partial \hat{t}}+\hat{v}_{i\alpha }\frac{%
\partial \hat{f}_{i}}{\partial \hat{x}_{\alpha }}=\frac{1}{\varepsilon \hat{%
\tau}}[\hat{f}_{i}^{eq}-\hat{f}_{i}]\text{,}  \label{SEC1-e9}
\end{equation}%
where the caret symbols are used to denote non-dimensional quantities $\hat{v%
}_{i\alpha }=v_{i\alpha }/e_{r}$, $\hat{t}=te_{r}/L$ , $\hat{\tau}=\tau
/t_{c}$, and $\hat{f}_{i}=f_{i}/\rho _{r}$. The parameter $\varepsilon
=t_{c}e_{r}/L$ is the Knudsen number which may be interpreted as either the
ratio of collision time to flow time or as the ratio of mean free path to
the characteristic flow length. We will not use the caret notation further
but will assume that the equation are in non-dimensional form henceforth.

In the two-dimensional case, the KT discrete velocity model has nine
components. It reads
\begin{equation}
(v_{i1},v_{i2})=\left\{
\begin{array}{ll}
& (0,0),i=0 \\
& c_{1}\left( \cos (\frac{\pi (i+1)}{2}),\sin (\frac{\pi (i+1)}{2})\right)
,i=1,2,3,4 \\
& c_{2}\left( \cos \pi (\frac{i+1}{2}+\frac{1}{4}),\cos \pi (\frac{i+1}{2}+%
\frac{1}{4})\right) ,i=5,6,7,8%
\end{array}%
\right.  \label{e14}
\end{equation}

\begin{equation}
\eta _i=\left\{
\begin{array}{ll}
\eta _0 & i=0 \\
0 & i=1,2,...,8%
\end{array}
\right. .  \label{e15}
\end{equation}

A schematic figure of the distribution of the discrete velocities is
shown in Fig.1, where $c_{1}$ and $c_{2}$ are constants which should
not depart faraway from the flow velocity $u$ and $c_{2}$ is generally chosen $%
1.0\sim 3.0$ times of $c_{1}$.
\begin{figure}
\includegraphics*[width=0.45\textwidth]{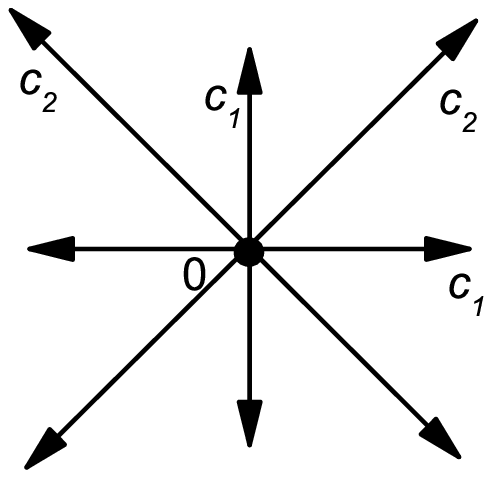}
\includegraphics*[width=0.45\textwidth]{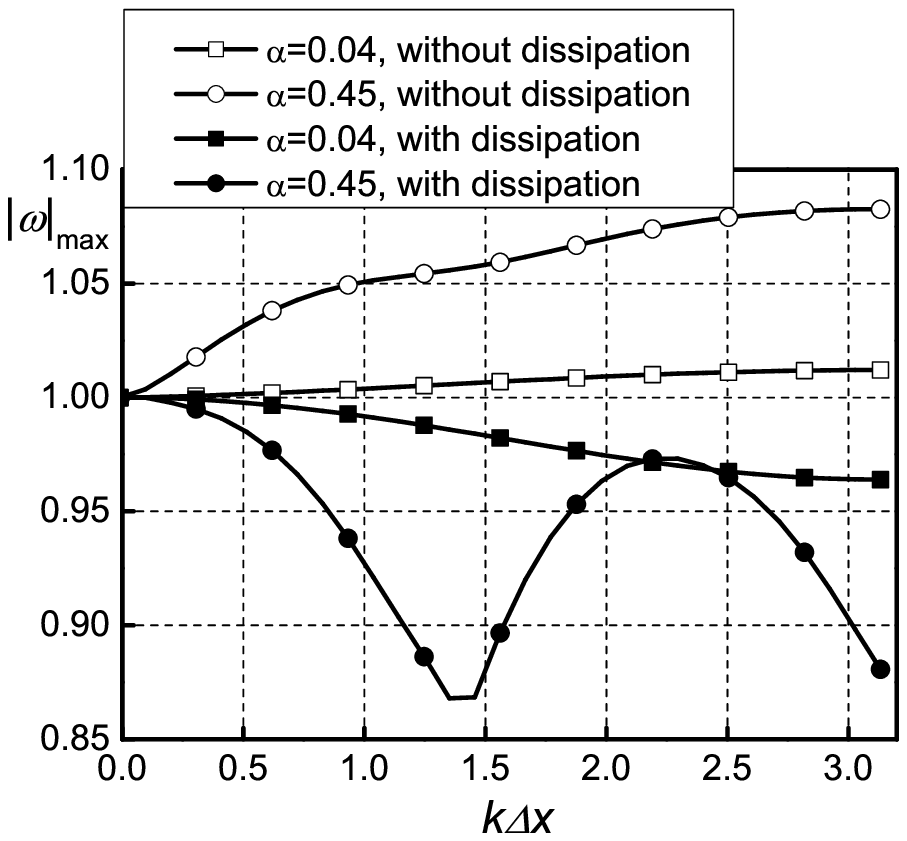}
\caption{(Left above)Schematic figure of the discrete velocity
model.}

\caption{(Right above) Effects of the dissipation term. Parameters used are $\protect\rho =1.0$, $%
T=1.0$, $u_{1}=10.0$, $u_{2}=0.0$, the remaining constants are set as $%
c_{1}=10$, $c_{2}=20$, $\protect\eta _{0}=10$, $\Delta t=\protect\alpha %
\Delta x/c_{2} $}
\end{figure}

The local equilibrium distribution function is computed by
\begin{equation}
f_{i}^{eq}=\rho (A_{i}+B_{i}v_{i\alpha }u_{\alpha }+D_{i}u_{\alpha
}v_{i\alpha }u_{\beta }v_{i\beta })\text{, }i=0,1,2,...,8\text{,}
\label{e16}
\end{equation}%
where
\begin{equation}
A_{i}=\left\{
\begin{array}{ll}
& \frac{b-2}{\eta _{0}}T\text{, }i=0 \\
& \frac{1}{4(c_{1}^{2}-c_{2}^{2})}\left[ -c_{2}^{2}+\left( (b-2)\frac{%
c_{2}^{2}}{\eta _{0}^{2}}+2\right) T+\frac{c_{2}^{2}}{c_{1}^{2}}u_{\alpha
}^{2}\right] \text{, }i=1,2,3,4 \\
& \frac{1}{4(c_{2}^{2}-c_{1}^{2})}\left[ -c_{1}^{2}+\left( (b-2)\frac{%
c_{1}^{2}}{\eta _{0}^{2}}+2\right) T+\frac{c_{1}^{2}}{c_{2}^{2}}u_{\alpha
}^{2}\right] \text{, }i=5,6,7,8%
\end{array}%
\right.  \label{e17}
\end{equation}

\begin{equation}
B_{i}=\left\{
\begin{array}{ll}
0,\quad i=0 &  \\
\frac{-c_{2}^{2}+(b+2)T+u_{\beta }^{2}}{2c_{1}^{2}(c_{1}^{2}-c_{2}^{2})}%
,\quad i=1,2,3,4 &  \\
\frac{-c_{1}^{2}+(b+2)T+u_{\beta }^{2}}{2c_{2}^{2}(c_{2}^{2}-c_{1}^{2})}%
,\quad i=5,6,7,8 &
\end{array}%
\right. \text{, }D_{i}=\left\{
\begin{array}{ll}
0,\quad i=0 &  \\
\frac{1}{2c_{1}^{4}},\quad i=1,2,3,4 &  \\
\frac{1}{2c_{2}^{4}},\quad i=5,6,7,8 &  \\
&
\end{array}%
\right.  \label{e18}
\end{equation}

It is clear that $\eta_0$, $c_1$ and $c_2$ are independent
parameters in this DVM and the value of $\eta_0$ influences the
discrete local equilibrium distribution function $f_{i}^{eq}$ via
the expansion coefficient $A_i$. The combination of the above DVM
and the general FD scheme with first-order forward in time and
second-order upwinding in space composes the original FDLB model by
KT. The FDLB by KT has been validated via the Riemann problem in
subsonic flows\cite{Kataoka_Euler}.  In a LB simulation the
discretization in time and space introduces unphysical waves, and
the collision term introduces a physical dissipation when the system
deviates from the local equilibrium. If the physical dissipation is
strong enough so that the unphysical oscillations are not to be
amplified in the simulation procedure, we will have no instability
problem. The original LB model by KT is not stable when the Mach
number $M$ exceeds $1$\cite{Kataoka_Euler},
 which shows that an additional dissipation term is needed in such cases.
 To make practical the LB
simulation to the supersonic flows, we propose an alternative FD
scheme in the following part of this section. The proposed FD scheme
will be combined with an additional dissipation term to overcome the
numerical instability problem in the next section.

We use the usual first-order forward scheme in time. Since all the
quantities are now non-dimensional, to simplify the following analysis, the
time step $\Delta t$ is set to be numerically equal to the Knudsen number $%
\varepsilon $. Thus, from Eq.\eqref{SEC1-e9} we have
\begin{equation}
f_{i}(\mathbf{x},t+\Delta t)-f_{i}(\mathbf{x},t)+v_{i\alpha }\frac{\partial
f_{i}(\mathbf{x},t)}{\partial x_{\alpha }}\Delta t=\frac{1}{\tau }\left[
f_{i}^{eq}(\mathbf{x},t)-f_{i}(\mathbf{x},t)\right] \text{.}
\label{SEC-VON-1}
\end{equation}%
In Eq.\eqref{SEC-VON-1} the spatial derivative $\partial f_{i}/\partial x$
can be calculated by
\begin{equation}
\text{If }v_{ix}\geq 0,\quad \frac{\partial {f_{i}}}{\partial x}=\frac{\beta
f_{i}(x+\Delta x,t)+(1-2\beta )f_{i}(x,t)-(1-\beta )f_{i}(x-\Delta x,t)}{%
\Delta x}\text{;}  \label{SEC-VON-2a}
\end{equation}%
\begin{equation}
\text{If }v_{ix}<0,\quad \frac{\partial {f_{i}}}{\partial x}=\frac{(1-\beta
)f_{i}(x+\Delta x,t)-(1-2\beta )f_{i}(x,t)-\beta f_{i}(x-\Delta x,t)}{\Delta
x}\text{.}  \label{SEC-VON-2b}
\end{equation}%
In Eqs.\eqref{SEC-VON-2a} and \eqref{SEC-VON-2b}, $0\leq \beta \leq 0.5$. If
$\beta $ takes zero, then they are not other than the first order upwind
scheme in space; if $\beta $ takes $0.5$, they recover to the general
central difference scheme. $\partial f_{i}/\partial y$ can be calculated in
a similar way. Actually, Eqs.\eqref{SEC-VON-2a} and \eqref{SEC-VON-2b} can
be rewritten as
\begin{eqnarray}
\text{If }v_{ix} &\geq &0,\quad \frac{\partial {f_{i}}}{\partial x}=\frac{%
f_{i}(x,t)-f_{i}(x-\Delta x,t)}{\Delta x}  \label{SEC-VON-2c} \\
&&+\frac{\beta \Delta x[f_{i}(x+\Delta x,t)+f_{i}(x-\Delta x,t)-2f_{i}(x,t)]%
}{{\Delta x}^{2}}\text{;}  \nonumber
\end{eqnarray}%
\begin{eqnarray}
\text{If }v_{ix} &<&0,\quad \frac{\partial {f_{i}}}{\partial x}=\frac{%
f_{i}(x+\Delta x,t)-f_{i}(x,t)}{\Delta x}  \label{SEC-VON-2d} \\
&&-\frac{\beta \Delta x[f_{i}(x+\Delta x,t)+f_{i}(x-\Delta x,t)-2f_{i}(x,t)]%
}{{\Delta x}^{2}}\text{.}  \nonumber
\end{eqnarray}%
The second terms in the right-hand-side of Eqs.\eqref{SEC-VON-2c} and %
\eqref{SEC-VON-2d} can be regarded as some kind of artificial
viscosities which are used to reduce some unphysical phenomena such
as wall-heating\cite{Ball1996}, but they are not enough to be
effectively improve the stability of LB simulation, which means
additional dissipation term is needed for a practical LB simulation.
In the following sections the parameter $\beta $ is chosen to be
$0.25$ if not particularly stated.

\section{von Neumann Analysis}

\label{SECTION-STABILITY}

The stability problem of LB has been addressed and attempted for
some years
\cite{lbe-1,Yong2003,Xiong2002,Tosi2006,Ansumali2002,Li2004,Sofonea2004,Brownlee2007}
. Among them, the the entropic LB method\cite%
{Tosi2006,Ansumali2002} tries to make the scheme to follow the $%
H$-theorem; The FIX-UP method\cite{Tosi2006,Li2004} is based on the
standard BGK scheme, uses a third order equilibrium distribution
function and a self-adapting updating parameter to avoid
negativeness of the mass distribution function. Flux limiter
techniques are used to enhance the stability of FDLB by Sofonea, et
al\cite{Sofonea2004}. Adding minimal dissipation locally to improve
stability is also suggested by Brownlee, et al\cite{Brownlee2007},
but there such an approach is not explicitly discussed. All the
above mentioned attempts are for low Mach number flows. In this
paper we focus mainly on high speed flows.

Following Seta, et al\cite{SetaJSP}, in this paper we resort to the
von Neumann stability analysis to compose a stable LB scheme where
the additional dissipation is effective and minimal. The following
analysis is based on the FD scheme shown in Eqs.\eqref{SEC-VON-2c}
and \eqref{SEC-VON-2d}. In the von Neumann analysis the solution of
finite-difference equation is written as the familiar Fourier
series, and the numerical stability is evaluated by the magnitude of
eigenvalues of an
amplification matrix. The small perturbation $\Delta f_{i}$ is defined as $%
f_{i}(\mathbf{x},t)=\Delta f_{i}(\mathbf{x},t)+\bar{f_{i}^{0}}$, where $\bar{%
f_{i}^{0}}$ is the global equilibrium distribution function and is a
constant which does not vary in space or time and depends only on the mean
density, velocity and temperature. From Eq. \eqref{SEC-VON-1} we can obtain
\begin{equation}
\Delta f_{i}(\mathbf{x},t+\Delta t)-\Delta f_{i}(\mathbf{x},t)+v_{i\alpha }%
\frac{\partial \Delta f_{i}}{\partial x_{\alpha }}\Delta t=\frac{\Delta f_{j}%
}{\tau }\left[ \frac{\partial f_{i}^{eq}}{\partial f_{j}}-1\right] \text{.}
\label{SEC-VON-4}
\end{equation}%
The perturbation part $\Delta f_{i}(\mathbf{x},t)$ may be written as series
of complex exponents, $\Delta f_{i}(\mathbf{x},t)=F_{i}^{t}\mathrm{exp}(i%
\mathbf{k\cdot x})$, where $F^{t}$ is an amplitude at grid point
$\mathbf{x}$ and time $t$, $i$ is an imaginary unit, and $k_{\alpha
}$ is the wave number of sine wave in the domain with the highest
resolution $1/\Delta x_{\alpha}$. From Eqs. \eqref{SEC-VON-4} we
obtain $F_{i}^{t+\Delta t}=G_{i,j}F_{j}^{t}$ ,
where $G_{ij}$ is a matrix being used to assess amplification rate of $%
F_{i}^{t}$ per time step $\Delta t$. If the maximum of the eigenvalues of
the amplification matrix satisfies the  condition, $\mathrm{max}%
|\omega |\leq 1$, for all wave numbers, the FD scheme is surely
stable, where $\omega $ is the eigenvalue of the amplification
matrix. This is the von Neumann condition for stability.

The amplification matrix $\mathbf{G}$ can be written as following,
\begin{equation}
G_{ij}=\left( 1-\frac{v_{i\alpha }\Delta t}{\Delta x_{\alpha }}\phi -\frac{1%
}{\tau }\right) \delta _{ij}+\frac{1}{\tau }\frac{\partial f_{i}^{eq}}{%
\partial f_{j}}  \label{SEC-VON-8}
\end{equation}%
where
\begin{equation}
\phi =\left\{
\begin{array}{ll}
\beta \mathrm{exp}(i k_{\alpha }\Delta x_{\alpha })+(1-2\beta
)-(1-\beta )\mathrm{exp}(-i k_{\alpha }\Delta x_{\alpha }), & \text{%
if }v_{i\alpha }\geq 0\text{;} \\
(1-\beta )\mathrm{exp}(i k_{\alpha }\Delta x_{\alpha })-(1-2\beta
)-\beta \mathrm{exp}(-i k_{\alpha }\Delta x_{\alpha }), & \text{if }%
v_{i\alpha }<0\text{.}%
\end{array}%
\right.   \label{SEC-VON-9}
\end{equation}

Several researchers have analyzed the stability of the
incompressible LB models\cite{SetaJSP,James1996,Niu2004}, it is
found that there is not a
single wave-number being always the most unstable. For the 2D DVM by KT $%
\mathbf{G}$ is a matrix with $9\times 9$ elements. Every element is
related to the macroscopical variables (density, temperature,
velocities), discrete velocities and other constants, so it is
difficult to analyze with explicit expressions. We resort to the
software, Mathematica-5.

In order to simulate high-speed flows, we introduce the following
dissipation term to the LB equation,
\begin{equation}
f_{i}(\mathbf{x},t+\Delta t)-f_{i}(\mathbf{x},t)+v_{i\alpha }\frac{\partial
f_{i}(\mathbf{x},t)}{\partial x_{\alpha }}\Delta t-\lambda _{i}\sum_{\alpha
=1}^{2}\frac{\partial ^{2}f_{i}(\mathbf{x},t)}{\partial x_{\alpha }^{2}}%
\Delta t=\frac{1}{\tau }[f_{i}^{eq}(\mathbf{x},t)-f_{i}(\mathbf{x},t)]
\label{SEC-VON-11}
\end{equation}%
where $\lambda _{i}$ is a small number not varying in space or time. The
second-order derivative $\frac{\partial ^{2}f_{i}(\mathbf{x},t)}{\partial
x_{\alpha }^{2}}$ can be calculated by the central difference scheme. Then $%
G_{i,j}$ can be written as
\begin{align}
G_{i,j} =\frac{\partial f_{i}^{eq}}{\partial f_{j}}-\frac{v_{i\alpha }\Delta t}{%
\Delta x_{\alpha }}\phi \delta _{ij}-\lambda _{i}\sum_{\alpha =1}^{2}\frac{%
2-2\mathrm{cos}(k_{\alpha }\Delta x_{\alpha })}{(\Delta x_{\alpha })^{2}}%
\Delta t\delta _{ij}\text{ .}  \label{SEC-VON-12}
\end{align}

Obviously, in Eq.\eqref{SEC-VON-12} the last term is required to
improve the numerical stability. How to chose the $\lambda _{i}$ is
the key problem here. It will not be effective if too small and will
result in too additional errors if too large. To get some indication
we look back to the last terms in Eqs.\eqref{SEC-VON-2c} and
\eqref{SEC-VON-2d} which are regarded as artificial viscosities to
reduce the numerical wall-heating phenomena. To simplify the
discussion, we choose always $\Delta x=\Delta y$. Indicative
analysis and numerical tests show that we can choose $\lambda _{i}$
around the following way,
\begin{equation}
\lambda _{i}=\left\{
\begin{array}{ll}
c_{1}\Delta x, & i=0 \\
c_{1}\Delta x/10, & i=1,2,3,4 \\
0, & i=5,6,7,8%
\end{array}%
\right. \text{.}  \label{SEC-VON-13}
\end{equation}

Now we show some results of von Neumann analysis by Mathematica-5 to get a
more complete understanding of the stability condition. We will show only
the results for high-Mach-number flows where the instability problem is
generally much more pronounced and previous LB models fail to work. The
results will be shown by figures with curves for the maximum eigenvalue $%
|\omega |_{max}$ of $\mathbf{G}$ versus $k\Delta x$. The wave number
$k$ is discretized from $0$ to $\pi $ with $30$ steps. Figure 2
shows a comparison between the two cases, with and without the
additional dissipation term, where the macroscopic variables are
chosen as $\rho=1.0,T=1.0,u_{1}=10.0,u_{2}=0.0$, and the constants in Eqs.\eqref{e14}
and %
\eqref{e15} are set as $c_{1}=10$, $c_{2}=20$, $\eta _{0}=10$. Coefficient $%
\alpha $ in the inset of the figure is a new constant introduced to
control the time step in the following way, $\Delta t=\alpha \Delta
x/c_{2}$.  For the two sets of results shown in the figure, it is
clear that the
dissipation term can significantly decrease the the maximum eigenvalue $%
|\omega |_{max}$ from being larger than to be smaller than $1$ for
appropriately given time step.

It is interesting to investigate the effects of various parameters
(physical quantities and model constants in Eqs.\eqref{e14} and
\eqref{e15}) on the numerical stability. Fig.3 shows the a
comparison of two cases: the first one is $\beta =0.25$ with
switching on the additional dissipation and the second is $\beta =0$
with switching off of the additional dissipation. The latter
corresponds to the conventional first-order upwind scheme. For the
given parameters, when the time step is small, both treatments give
stable simulations; but when the time step becomes large, the first
treatment makes the simulation stable while the second one does
not.

\begin{figure}[tbp]
\includegraphics*[width=0.47\textwidth]{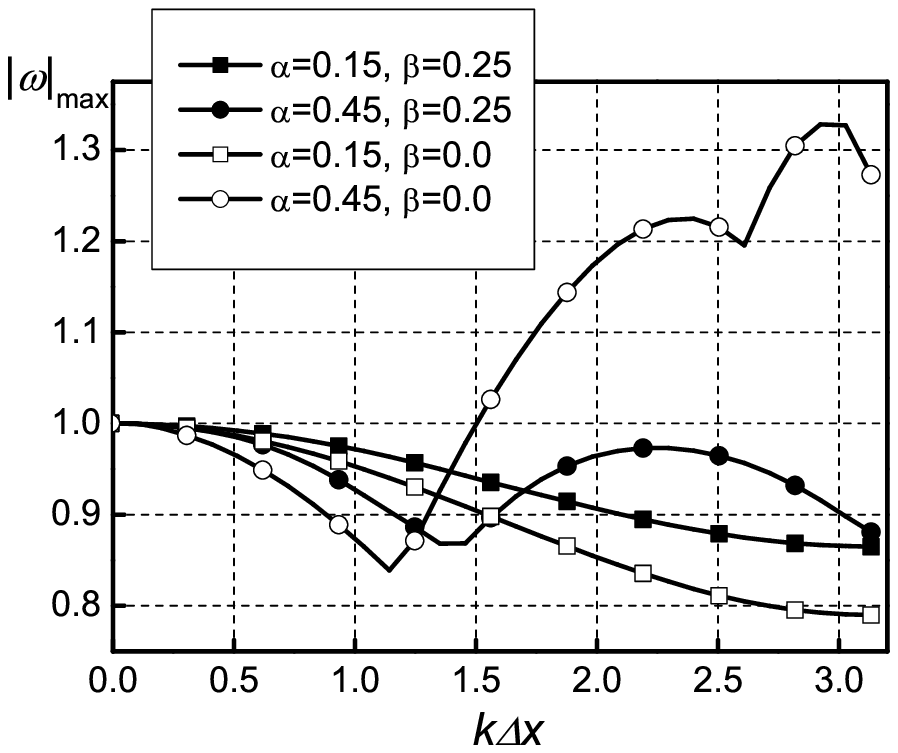}
\includegraphics*[width=0.47\textwidth]{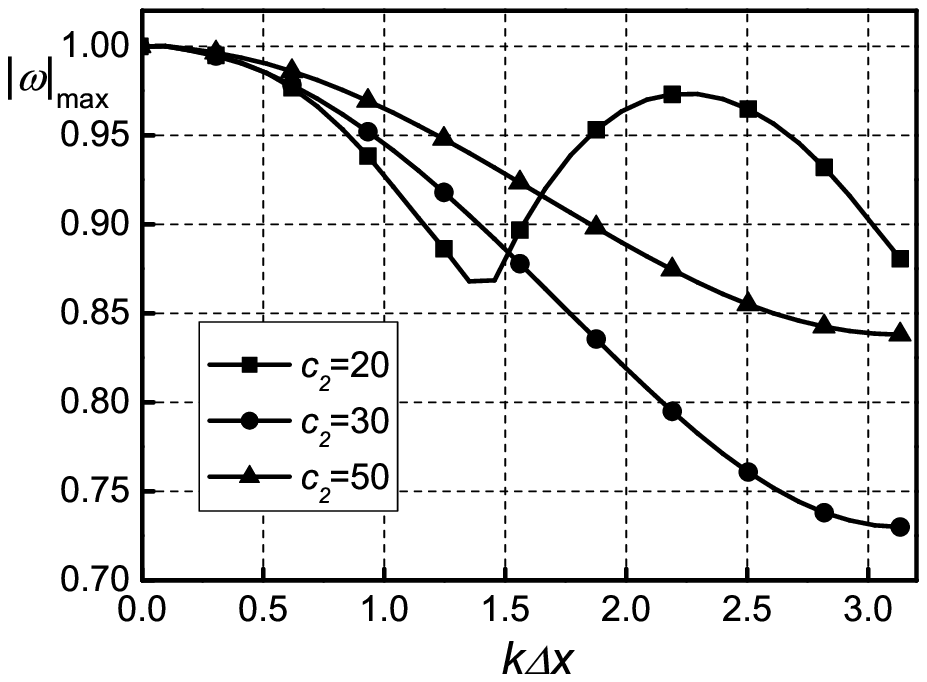}
\caption{(Left above) Stability analysis for mixed schemes. The
macroscopic variables are set as $\protect\rho
=1.0,T=1.0,u_{1}=10.0,u_{2}=0.0$, the constants are set as
$c_{1}=10,c_{2}=20,\protect\eta _{0}=10,\Delta t=\protect\alpha
\Delta x/c_{2}$.}

\caption{(Right above) The influence of $c_{2}$. The macroscopic variables are set as $%
\protect\rho =1.0,T=1.0,u_{1}=10.0,u_{2}=0.0$, the other constants
are set as $c_{1}=10,\protect\eta _{0}=10,\Delta t=0.45\Delta
x/c_{2}$.}
\end{figure}

Figure 4 shows an investigation to the influence of constant $c_{2}$
on the stability of LB simulation, where the value of $c_{2}$ is
altered from $10$ to $50$, the time step $\Delta t=0.45\Delta
x/c_{2}$, the other constants and macroscopic variables are
unchanged. The LB is stable for all tested values of $c_{2}$. Our
experience shows that the value of $c_{2}$
does not influence much the numerical stability if it is not smaller than $%
2c_{1}$, but the stable time step becomes smaller for larger the value of $%
c_{2}$.
Figure 5 shows an investigation to the influence of the value of $%
\eta _{0}$. The value of $\eta _{0}$ is altered from $5$ to $20$, $c_{2}=20$%
, $\Delta t=0.3\Delta x/c_{2}$, the other constants and macroscopic
variables are kept unchanged. We get an indication that it is not difficult
to find an appropriate value of $\eta _{0}$ to get a stable simulation. For
cases shown in the figure, only a too small value of $\eta _{0}$ may result
in instability (see the case of $\eta _{0}=5$) and stability is nearly the
same when $\eta _{0}$ exceeds some critical value (see the cases with $\eta
_{0}=15$ and with $\eta _{0}=20$).
\begin{figure}[tbp]
\includegraphics*[width=0.47\textwidth]{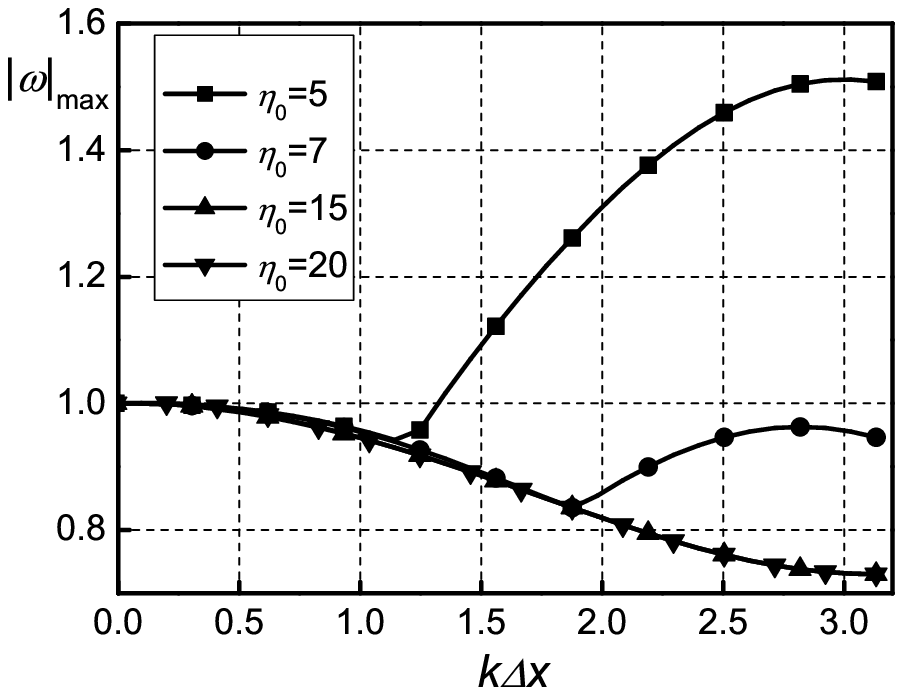}
\includegraphics*[width=0.47\textwidth]{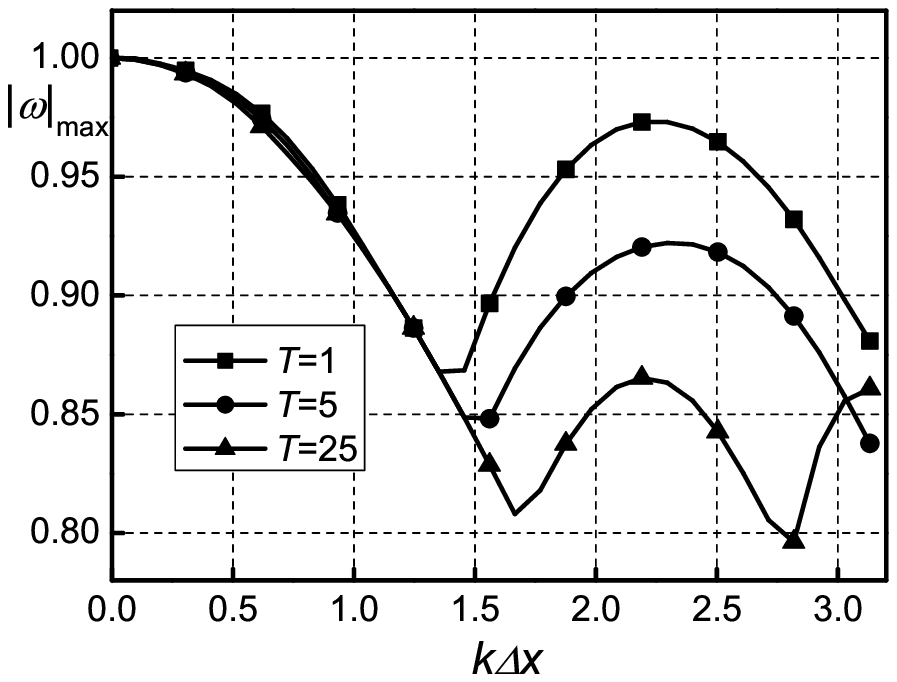}
\caption{(Left above) The influence of $\protect\eta _{0}$. The
three physical quanties are set as $\protect\rho
=1.0,T=1.0,u_{1}=10.0,u_{2}=0.0$, and other constants are set as
$c_{1}=10,c_{2}=20,\protect\eta _{0}=10,\Delta t=0.3\Delta
x/c_{2}$.}

\caption{(Right above) The influence of temperature $T$. The other
physical quantities are set as $\protect\rho
=1.0,u_{1}=10.0,u_{2}=0.0$, the constants are set as
$c_{1}=10,c_{2}=20,\protect\eta _{0}=10,\Delta t=0.45\Delta
x/c_{2}$.} \label{Fig6}
\end{figure}

Since the density $\rho $ can be normalized to $1$, we then
investigate only the effects of the other two physical quantities,
temperature $T$ and flow velocity $\mathbf{u}$. Figure 6 shows three
cases with different temperatures, $T=1$, $T=5$ and $T=25$. When
other parameters are fixed, the numerical stability increase with
the increasing of the system temperature. This can also be
understood that higher temperature corresponds to higher sound speed
and lower Mach number.

Figure 7 shows cases with difference flow velocities. The value of $%
u_{1}$ is altered from zero to $15$ and $u_{2}=0$. For parameters used in
this case, we can find that the simulation will not be stable if $u_{1}$ is
much larger than $c_{1}$, even though $|\omega |_{max}$ is only slightly
larger than $1$ at $k\Delta x\approx 0.5$. Our experience shows that the
value of $c_{1}$ can be set nearly equal to the maximum of the flow
velocity.
\begin{figure}
\includegraphics*[width=0.47\textwidth]{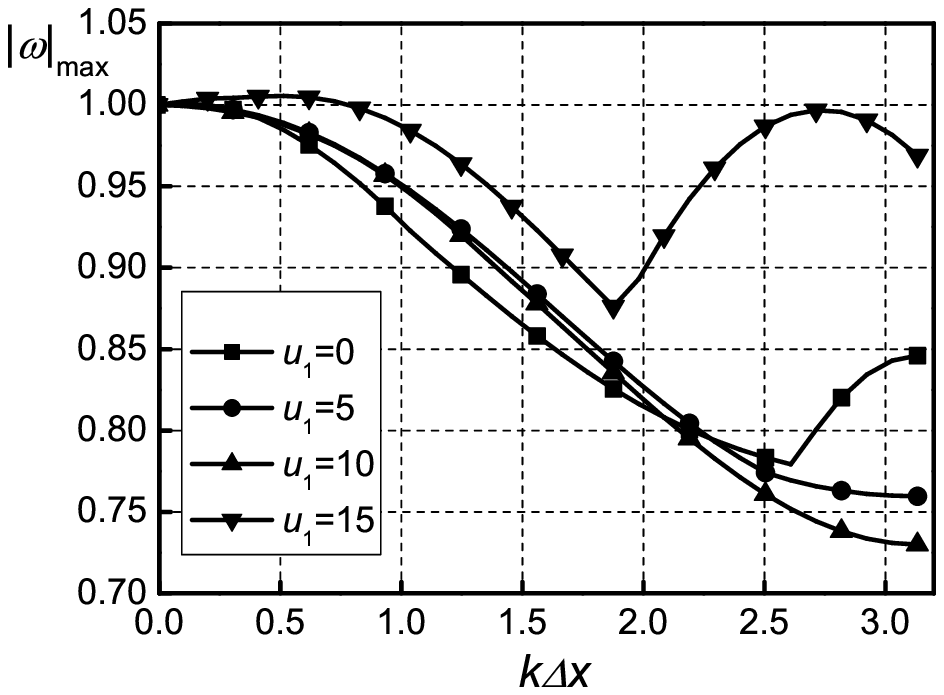}
\includegraphics*[width=0.47\textwidth]{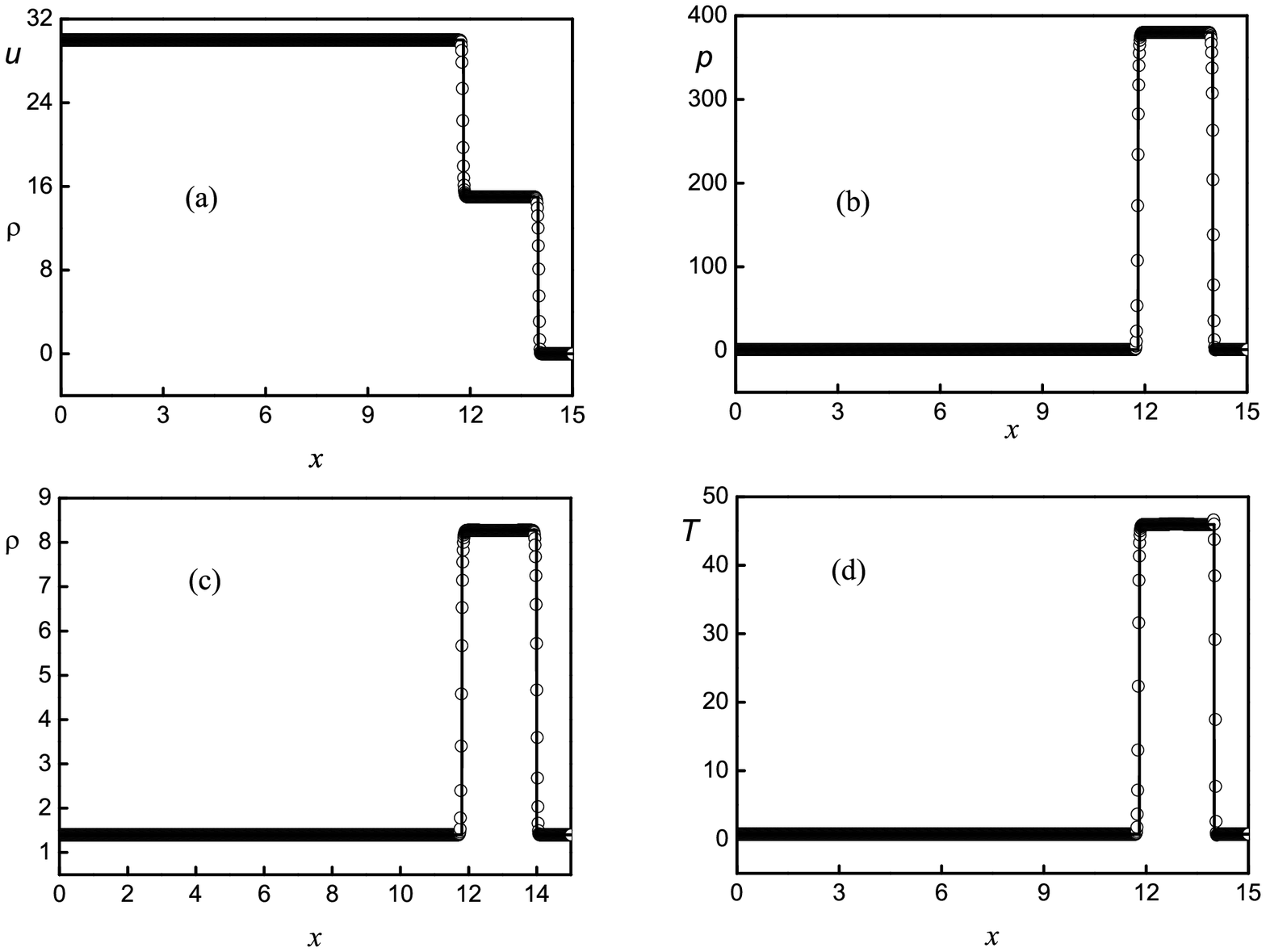}
\caption{(Left above) The influence of flow velocity $u_{1}$. The
other physical quantities are set as $\protect\rho
=1.0,T=1.0,u_{2}=0.0$, the constants are set as
$c_{1}=10,c_{2}=20,\protect\eta _{0}=10,\Delta t=0.3\Delta x/c_{2}$.
}

\caption{(Right above)The $x$ dependence of $\protect\rho$, $p$, $u$
and $T$. The symbols are simulation results by the new LB and lines
are analytic solutions. The initial condition is described by Eq.
\eqref{Eq-Riemann-1}. Here $\Delta x=0.01$, $\Delta t=0.00008$, $c_1=25$,$%
c_2=50$,$\protect\eta_0=30$, terminal time $t=0.36$. }
\end{figure}

In summary, constants $c_1$, $c_2$ and $\protect\eta_0$ influence
heavily the stability. In practical simulations, $c_{1}$ can be set
 approximately equal to the maximum of flow velocity; $c_{2}$ can be
set to be about $2\sim 3$ times of the value of $c_{1}$; $\eta _{0}$
can be set an appropriate value in between $c_{1}$ and $c_{2}$.
Equation \eqref{SEC-VON-13} is indicative in choosing parameters for
stable LB simulations of high-speed flow.

\section{Numerical validations}

Two kinds of benchmarks are used to validate the proposed scheme. The first
one is the Riemann problem. The second is the problem of shock reflection.

\subsection{Riemann problem}

Here the two-dimensional model is used to solve the one-dimensional Riemann
problem. The initial macroscopic variables at the two sides are $\rho _{L}$,
$p_{L}$ and $u_{L}$, and $\rho _{R}$, $p_{R}$ and $u_{R}$, respectively. We
firstly simulate a Riemann problem with an initial condition described by
\begin{equation}
\begin{array}{ll}
\rho _{L}=1.4, & \rho _{R}=1.4, \\
p_{L}=1.0, & p_{R}=1.0, \\
u_{L}=30.0, & u_{R}=0.0,%
\end{array}
\label{Eq-Riemann-1}
\end{equation}%
where the subscripts``L" and ``R" denote the left and right sides of
the discontinuity. The initial Mach number of left flow is equal to
$30.0$. The numerical results for $\gamma =1.4$ are shown in Fig.8,
where the symbols are simulation results and solid lines are
analytical solutions. The
parameters used in the simulation are $c_{1}=25$, $c_{2}=50$, $\eta _{0}=30$%
, $t=0.36$. The size of grid is $\Delta x=\Delta y=0.01$. Time step
$\Delta t=0.00008$. The two sets of results have satisfying
agreement. In this case no evident \textquotedblleft
wall-heating\textquotedblright\ phenomenon is observed. As a
comparison, we show a result with the general first order upwind
scheme for the pressure in Fig.9(a). A abrupt decrease in pressure
around $x=12$ corresponds to  the well known wall-heating
phenomenon. In order to observe the effects of various additional
viscosity, we vary the
value of $\lambda_0$ from $c_1 \Delta x$ to $5c_1 \Delta x$ under the fixed $%
\lambda_i(i=1,...,8)$. Figure 9 (b) shows the simulation results and
the exact one. We can find that the numerical width of shock becomes
wider and wall-heating problem becomes more pronounced as
$\lambda_0$ increases. Results in Fig. 9 confirm that
\eqref{SEC-VON-13} is indicative in choosing the additional
viscosity.

\begin{figure}[tbp]
\centering
\includegraphics*[width=0.95\textwidth]{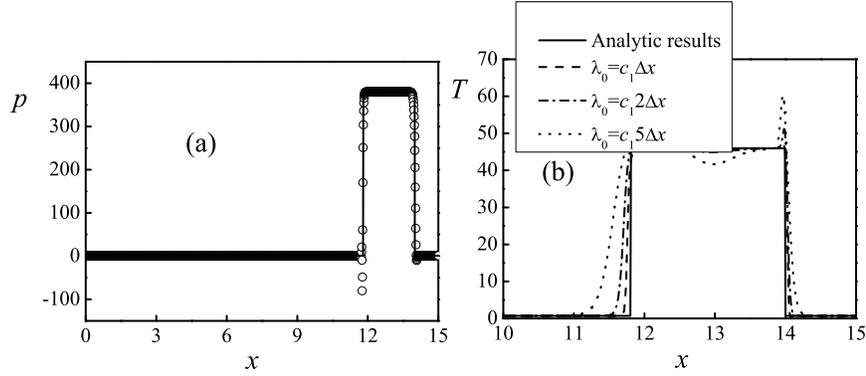}
\caption{Comparison of various finite-difference schemes (a) and
artificial viscosities (b). The initial
condition is same as Fig.8. Here $\Delta x=0.01$, $%
\Delta t=0.00008$, $c_1=25$,$c_2=50$,$\protect\eta_0=30$,  $%
t=0.36$.  (a) Profile of pressure. The values of
$\protect\lambda_i(i=0,...,8)$ are the same as those in Fig.8, while
$\beta=0$. The symbols correspond to simulation result the line is
for analytic solution. (b) Profile of temperature. The dashed, dash
dotted and dotted lines correspond to $\protect\lambda_0= c_1 \Delta
x$, $2c_1 \Delta x$, and $5c_1 \Delta x$, respectively. The values
of $\protect\lambda_i(i=1,...,8)$ and $\beta$ are the same as in
Fig.8. }
\end{figure}

The second example is the propagation of a shock with high ratios of density
and pressure. The initial macroscopic variables are give by
\begin{equation}
\begin{array}{ll}
\rho _{L}=1000.0, & \rho _{R}=1.0, \\
p_{L}=1000.0, & p_{R}=1.0, \\
u_{L}=0.0, & u_{R}=0.0,%
\end{array}
\label{Eq-Riemann-2}
\end{equation}%
The size of grid is $\Delta x=\Delta y=2.5\times 10^{-3}$. The
numerical results are shown for $\gamma =1.4$ in Fig.10, where the
symbols are simulation results and solid lines correspond to exact
solutions. We find also a good agreement between the two sets of
results.

\begin{figure}
\centering
\includegraphics*[width=0.7\textwidth]{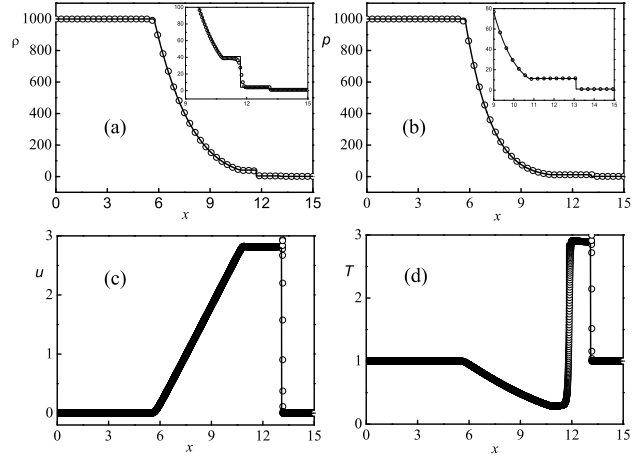}
\caption{The $x$ dependence of $\protect\rho$, $p$, $u$ and $T$. The symbols
are simulation results by the new LB and lines are analytic solutions. The
initial condition is described by Eq. \eqref{Eq-Riemann-2}. The used
parameters are $\Delta x=0.0025$, $\Delta t=0.0002$, $c_1=3$,$c_2=9$,$%
\protect\eta_0=5$, terminal time $t=1.5$. }
\end{figure}

\subsection{Shock reflection}

We will present two gas dynamics simulations. Both are done on
rectangular grid. The first is to recover a steady regular shock
reflection. The second test problem is the double Mach reflection of
a shock off an oblique surface. This example is used in Ref.
\cite{Woodward1984} as a benchmark test for comparing the
performance of various difference methods on problem involving
strong shocks.

In the first test problem, we have performed a $30^{\circ }$ shock
reflection for $\gamma =1.4$. The computational domain is a rectangle with
length 9 and height 3 (See Fig.11(a)). This domain is divided into a $%
900\times 300$ rectangular grid with $\Delta x=\Delta y=0.01$. The
boundary conditions are composed of a reflecting surface along the
bottom boundary, supersonic outflow along the right boundary, and
Dirichlet conditions on the other two sides, given by
\begin{equation}
\begin{array}{l}
(\rho ,u_{1},u_{2},p)|_{0,y,t}=(1.0,10.0,0.0,1/1.4) \\
(\rho ,u_{1},u_{2},p)|_{x,1,t}=(5.0,8.0,-3.4641,20.7143)%
\end{array}
\label{Eq-Riemann-3}
\end{equation}%
Initially, we set the solution in the entire domain to be that at
the left boundary, the corresponding Mach number is $10.0$. In
Fig.11(b) we show a contour plot of the density. The clear shock
reflection on the wall agrees well with the exact solution.

\begin{figure}[tbp]
\caption{(See 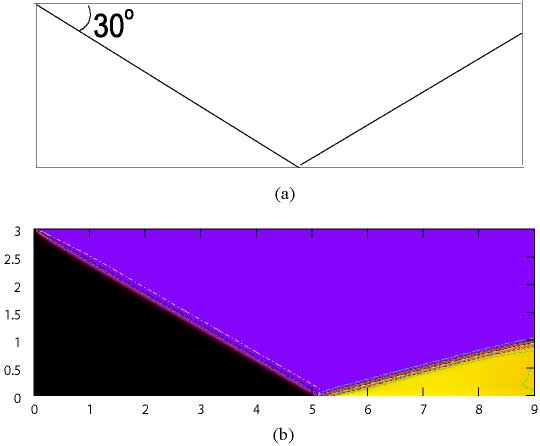 )(Color online) Regular shock reflection.
$(a)$, Sketch map of the
steady state regular reflection problem. $(b)$, The density contour at time $%
t=2.5$ with $\Delta x = \Delta y = 0.01$, $\Delta t = 1.5\times10^{-4}$, $%
c_1=10$,$c_2=20$,$\protect\eta_0=15$; Left and up boundary
conditions are given by Eq.\eqref{Eq-Riemann-3}.  From black to
yellow, the value increases.}

\caption{(See 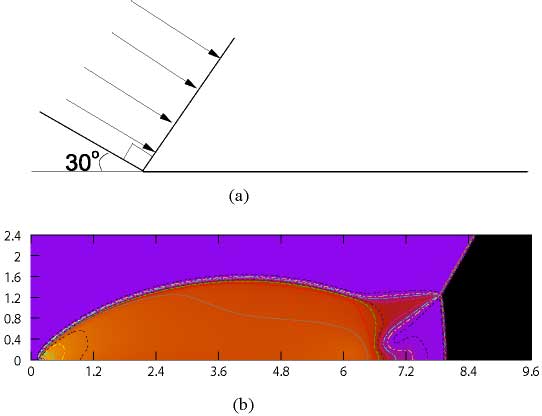) (Color online) Double Mach reflection.
$(a)$ initial configuration; $(b)$ the density contour at time
$t=1.5$ with $\Delta x = \Delta y = 0.01$, $\Delta t = 2.0\times
10^{-4}$, $c_1=8$,$c_2=16$,$\protect\eta_0=10$. The reflecting wall
begins at $20$ mesh length from the lower left corner. From black to
yellow, the value increases.}
\end{figure}

The second test problem is an unsteady shock reflection. A planar
shock is incident on an oblique surface with the surface at a
$30^{\circ }$ angle to the direction of propagation of the shock
(Fig.12(a)). The fluid
in front of the shock has zero velocity, and the shock Mach number is $10.0$%
. In Fig.12(b) we show the result of density contour, where the
double Mach reflection phenomenon is successfully recovered.

\section{Conclusions and discussions}

The lattice Boltzmann simulation to high-speed compressible flows is
revisited by proposing an improved LB model. The new LB model is
composed of the original discrete-velocity-model by Kataoka and
Tsutahara and an appropriate finite-difference scheme to the
convection term. An additional dissipation term is introduced to
improve the numerical stability. The adding of the dissipation term
should survive the dilemma of stability versus accuracy. In other
words, the dissipation should be minimal but make the evolution
satisfy the von Neumann stability condition. The effects of
polynomial equilibria\cite{Yong2003} are taken into account (via the
first term of Eq.\eqref{SEC-VON-12}) in such an approach. Due to the
complexity the analysis resorts to the software, Mathematica-5, and
only some typical results are shown by figures.

Benchmark tests are used to validate the proposed scheme and
reference values of model parameters are suggested. Typical Riemann
problems with high-Mach-number ($30$ or higher) and high ratios
($1000:1$) of pressure and density show good accuracy and stability
of the new scheme, even though they are generally difficult to
resolve by traditional computational fluid dynamics. Regular and
Mach shock reflection problems are successfully recovered, which
shows also the potential application of lattice Boltzmann model to
fluid systems where non-equilibrium processes are intrinsic and
pronounced. The new LB model may be used to investigate some
long-standing problem, such as the transition between regular and
shock reflections. At the moment, we are still not able to present a
complete description on the most appropriate additional dissipation
term, but the idea presented in the paper can be easily used to get
some practically useful solutions for stability enhancement. We plan
to better clarify the physical dissipation and artificial ones in
the future.

\section*{acknowledgments}

We warmly thank Profs. Jianshi Zhu, Xingping Liu, Xijun Yu, Zhijun
Shen, and Yingjun Li for helpful discussions. Suggestions from the
anonymous referee are gratefully acknowledged. This work is partly
supported by the National Basic Research Program [Grant No.
2005CB321700], National Natural Science Foundation  [Grant No.
10474137] of China, and Science Foundation of Laboratory of
Computational Physics, Institute of Applied Physics and
Computational Mathematics, Beijing, China.

\end{document}